%% file: main.tex
\documentclass{jps-cp}
\makeatletter
\renewcommand{\@biblabel}[1]{[#1]}
\makeatother

\usepackage{amssymb}
\usepackage{amsfonts}
\usepackage{xcolor}

\newcommand{\orcid}[1]{\href{https://orcid.org/#1}{\textsuperscript{[ORCID]}}}
\usepackage{amsmath}
\usepackage{comment}
\usepackage{braket}
\usepackage{booktabs}

\usepackage{graphicx}
\input{macros} 

\title{An Exact Conjugation Identity for the Many-Body Wilson Loop Beyond Quantization}%

\author{Kai Watanabe}
\inst{}
\recdate{\today}

\abst{
Constraints on unquantized many-body holonomies, such as Wilson loops (or their Berry phases), are less explored than those on their quantized counterparts.
Here, we realize an unquantized regime by tuning the bond dimerization $\delta$ and the staggered potential $\Delta$ in a half-filled dimerized staggered Hubbard ring.
For the tuned parameter sets, a finite excitation gap persists along the $U(1)$ twist cycle $\theta\in[0,2\pi]$, so that the ground state $|\psi_{\delta}(\theta)\rangle$ remains separated from the excited states.
The many-body Wilson loop is therefore well defined from the ground-state family $\{|\psi_{\delta}(\theta)\rangle;\,\theta\in[0,2\pi]\}$.
In this setup, we show an exact many-body Wilson loop conjugation identity,
$W(-\delta)=W(\delta)^*$,
accumulated along a cycle parametrized by $\theta$.
Importantly, the identity persists in regimes where the Berry phase $\gamma\equiv-\arg W$ varies continuously.
We demonstrate the identity numerically using the density-matrix renormalization group (DMRG) method.
The identity extends to other models where the flux-threaded ground-state family along the closed $\theta$-cycle is mapped to the reversed cycle.
Beyond its conceptual content, the identity provides a symmetry-based consistency check for numerical evaluations of Berry phases in interacting systems.
It also justifies the signal-to-noise ratio improvement in Monte Carlo simulations by performing simulations at both $\delta$ and $-\delta$ and averaging $W(\delta)$ with $W(-\delta)^{*}$.
}
\kword{Berry phase, Topology, Hubbard model, Wilson loop, Gauge theory}
\begin{document}
\maketitle

\input{sections/01_intro}
\input{sections/02_model}
\input{sections/04_results}
\section*{Acknowledgements}
The author thanks Akito Daido and David Vanderbilt for helpful discussions, and the Yukawa Institute for Theoretical Physics at Kyoto University for hosting the YITP workshop YITP-W-26-07, where discussions contributed to this work.
\bibliographystyle{plain}
\bibliography{refs}

\end{document}

%% file: macros.tex
\providecommand{\Eq}[1]{Eq.~(\ref{#1})}

\providecommand{\Fig}[1]{Fig.~\ref{#1}}

%% file: sections/01_intro.tex
\section{Introduction}
The many-body Wilson loop, defined from overlaps of an eigenstate family along a $U(1)$ flux-threading cycle, provides a compact characterization of topological many-body systems.
It is directly related to physical observables such as Berry phase, polarization and pumped charge~\cite{Zak1989,KingSmithVanderbilt1993,Resta1998}.
Most existing discussions focus on symmetry-enforced quantized regimes, where the observables can be interpreted in terms of topological invariants such as Chern numbers and winding numbers~\cite{ChiuTeoSchnyderRyu2016}.

However, useful symmetry constraints on many-body Wilson loops also arise in unquantized regimes.
For example, Shiozaki gave a group-theoretical treatment that describes how Wilson loops transform under continuous contour deformations without assuming quantization~\cite{Shiozaki2026Equivariant}.

In this work, we show an exact Wilson loop identity in an interacting lattice model.
In particular, in a dimerized lattice model with a dimerization $\delta$, the exact conjugation identity reads
\begin{equation}
\label{eq:W_conj_intro}
W(-\delta)=W(\delta)^*.
\end{equation}
As a corollary, the Berry phase $\gamma(\delta)\equiv -\arg W(\delta)$ satisfies $\gamma(-\delta)=-\gamma(\delta)\ (\mathrm{mod}\ 2\pi)$.

Crucially, Eq.~(\ref{eq:W_conj_intro}) holds without quantization of $\gamma(\delta)$. (We demonstrate this explicitly in gapped but unquantized regimes realized by tuning a Rice--Mele--type staggered potential together with the bond dimerization in later sections.)
Accordingly, Eq.~(\ref{eq:W_conj_intro}) should be viewed as a Wilson-loop-level constraint beyond quantization.
Equivalently, Eq.~(\ref{eq:W_conj_intro}) fixes the even/odd parity of $\mathrm{Re}\,W(\delta)$ and $\mathrm{Im}\,W(\delta)$ under $\delta\to-\delta$, providing a direct and gauge-invariant signature of the composite symmetry.

Our result has two immediate implications:
(i) it provides a compact organizing principle for the $\delta\to-\delta$ dependence of the interacting system, and
(ii) it serves as a stringent diagnostic for twist-based holonomy computations.

While we mainly consider the dimerized model, the same technique extends to models in which the permutations of the hopping parameters are equivalent to reversing the bond pattern.

%% file: sections/02_model.tex
\section{Concrete example from a dimerized staggered Hubbard model}
\label{sec:model}
We consider a $U(1)$ flux $\theta$-threaded dimerized staggered Hubbard ring with even system size $L$ at half filling~\cite{RiceMele1982,Su1979SolitonsPolyacetylene}.
The Hamiltonian reads $\hat H(\delta,\theta)=\hat H_t(\delta,\theta)+\hat H_\Delta+\hat H_{\mathrm{corr}}$,
where
\begin{equation}
\label{eq:H_hop}
\hat H_t(\delta,\theta)=\sum_{j=1}^{L}\sum_{\sigma}
\left[\left\{1+(-1)^j\delta\right\}\,e^{i\theta/L}\,c^\dagger_{j\sigma}c_{(j+1)\sigma}+ \mathrm{H.c.}\right],
\end{equation}
\begin{equation}
\label{eq:H_stag}
\hat H_\Delta=\Delta\sum_{i=1}^{L}(-1)^i n_i,
\qquad
\left(n_i=\sum_{\sigma}n_{i\sigma}\right),
\end{equation}
and
\begin{equation}
\label{eq:H_U}
\hat H_{\mathrm{corr}}=U\sum_{i=1}^{L} n_{i\uparrow}n_{i\downarrow}.
\end{equation}
Here $c_{i\sigma}$ annihilates a fermion with spin $\sigma$ on site $i$.
The number operator reads $n_{i\sigma}=c^\dagger_{i\sigma}c_{i\sigma}$.
Periodic boundary condition (PBC) $c_{(L+1)\sigma}\equiv c_{1\sigma}$ is imposed in this work.
For numerical convenience, we employ a parametrization $\Delta=U/2-\lambda$.
The twist $\theta\in[0,2\pi]$ is introduced in a uniform gauge, i.e., the same $e^{i\theta/L}$ is attached to every nearest-neighbor hopping.
Throughout we work in the fixed-number sector
$N_\uparrow=N_\downarrow=L/2$.
In this setting, reversal of the bond dimerization $\delta\to-\delta$ implements the bond-pattern reversal by interchanging even and odd links.
We denote by $|\Psi_\delta(\theta)\rangle$ the many-body ground state of $\hat H(\delta,\theta)$ in this sector.

This model admits regimes with continuously varying $\gamma(\delta)$ at finite $\delta$ while the excitation gap remains finite. This is because, by varying $\delta$ and $\Delta$ simultaneously, one can change the phase without crossing $\delta=\Delta=0$ and without closing the gap~\cite{MorimotoFurusaki2013}.

We now show that there exists a composite antiunitary mapping $\Xi$ which maps the flux-threaded ground-state family along the $\theta$ cycle to the reversed family: 
\begin{equation}
  \label{eq:Xi_maps_psi}
|\Psi_{-\delta}(\theta)\rangle = e^{i\phi(\theta)}\,\Xi|\Psi_{\delta}(\theta)\rangle,
\end{equation}
where $\phi(\theta)$ is a physically irrelevant $\theta$-dependent phase that cancels on the closed $\theta$-cycle, as we show below.
Since $|\Psi_{\pm\delta}(\theta)\rangle$ are eigenstates of $\hat H(\pm\delta,\theta)$,
it suffices to establish
\begin{equation}
\label{eq:Xi_maps_H}
\Xi\,\hat H(\delta,\theta)\,\Xi^{-1}=\hat H(-\delta,\theta).
\end{equation}

For this model, $\Xi$ can be constructed as the product of a one-site translation $T_1$,
a particle-hole transformation ${\cal C}_{\rm ph}$, and a complex conjugation ${\cal K}$:
\begin{equation}
\label{eq:Xi_def}
\Xi \equiv {\cal K}\,{\cal C}_{\rm ph}\,T_1.
\end{equation}

The one-site translation acts as
$T_1\,c_{i\sigma}\,T_1^{-1}=c_{(i+1)\sigma}$ and $T_1\,n_{i\sigma}\,T_1^{-1}=n_{(i+1)\sigma}$,
where the PBC ensures the well definedness.
The particle-hole transformation reads ${\cal C}_{\rm ph}\,c_{i\sigma}{\cal C}_{\rm ph}^{-1}=(-1)^i c^\dagger_{i\sigma}$.
Simple algebra yields ${\cal C}_{\rm ph}\,n_{i\sigma}\,{\cal C}_{\rm ph}^{-1}=1-n_{i\sigma}$.

The composite mapping $\Xi$ acts on Hamiltonian $\hat H(\delta,\theta)$ as follows.
First, under $T_1$ the signs on the staggered factor and the dimerized bond flip as $(-1)^i\mapsto-(-1)^i$.
Thus, $\{1+(-1)^i\delta\}\mapsto\{1-(-1)^i\delta\}$ which is equivalent to $\delta\mapsto-\delta$.
Similarly, $\hat H_\Delta$ changes sign under $T_1$.
Second, ${\cal C}_{\rm ph}$ flips the sign of $\hat H_\Delta$ again,
because $n_i\mapsto 2-n_i$ and $\sum_i^{L={\mathrm{even}}}(-1)^i=0$, so that
${\cal C}_{\rm ph}\,\hat H_\Delta\,{\cal C}_{\rm ph}^{-1}=-\hat H_\Delta$.
The on-site correlation term $\hat H_{\mathrm{corr}}=U\sum_i n_{i\uparrow}n_{i\downarrow}$ is invariant
under ${\cal C}_{\rm ph}$ up to an additive constant proportional to $L-N_{\rm tot}$, which vanishes at half filling.
The particle-hole transformation also reverses the hopping orientation, thereby sending $\theta\to-\theta$ in the hopping term.
Finally, complex conjugation ${\cal K}$ flips the U(1) flux as
$e^{+i\theta/L}\mapsto e^{-i\theta/L}$, i.e., $\theta\mapsto-\theta$.
Combining these actions, we obtain the mapping in
\Eq{eq:Xi_maps_H}.

The many-body Wilson-loop is defined as
$
W(\delta)=\prod_{j=0}^{N_\theta-1}\mathcal{O}_{j}(\delta;\theta),
$
where
$\mathcal{O}_{j}(\delta;\theta)=\langle\Psi_\delta(\theta_j)\mid\Psi_\delta(\theta_{j+1})\rangle$ is the overlap.
The $\theta$-grid is defined as
$\theta_j=\frac{2\pi j}{N_\theta}$
and
$\theta_{N_\theta}\equiv\theta_0$.
\Eq{eq:Xi_maps_psi} implies that the overlaps transform by complex conjugation,
$\mathcal{O}_j(-\delta;\theta) = e^{i\phi(\theta_{j+1})-i\phi(\theta_j)}\mathcal{O}^*_{j}(\delta;\theta)$.
Taking the product over $j$ along the closed twist cycle $\theta_{N_\theta}\equiv\theta_0$, the additional phases cancel.
Hence, the Wilson-loop obeys the exact conjugation identity
$
W(-\delta)=W(\delta)^{*}.
$

%% file: sections/04_results.tex
\section{Numerical demonstration}
\label{Sec:result}
In this work, the ground states $|\Psi_\delta(\theta)\rangle$ are computed by the density-matrix renormalization group (DMRG)~\cite{White1992,HauschildPollmann2018TeNPy}.
We work on a ring of $L=12$ sites in the half-filled sector $N_{\uparrow}=N_{\downarrow}=L/2$.

In practice, we set the parameters $(U,\lambda,\delta)$ to a region where the charge and excitation gaps remain finite, though small.
In this regime, the two-parameter deformation space spanned by $(\delta,\lambda)$ allows gapped continuous deformations, so that $\gamma(\delta)$ is not pinned and can vary continuously.
Unless otherwise stated, we fix $\lambda=0.67$ and focus on two representative interaction strengths, $U=6.0$ and $U=7.0$.
We set the dimerization to $\delta=\pm0.005$ and $\pm0.01$.
With these parameters, the Berry phase is continuous (depinned).
We also include $\delta=0$ as a pinned reference case.

Throughout the parameter sets used in this study, the excitation gap
$\Delta E(\theta_j)=E_1(\theta_j)-E_0(\theta_j)$
attains its minimum at $j=0$ and remains finite there.
From this observation, we conclude that the gap is bounded from below, $\min_j \Delta E(\theta_j) > 7.0\times 10^{-2}$.
Moreover, the low-lying energies $E_n(\theta_j)$ show negligible dependence on the discretization $N_\theta$ over the resolutions used in our calculations.

We assess the reliability of the overlap Wilson loop $W$ by explicitly tracing the link overlaps $\{\mathcal{O}_j(\delta;\theta)\}$ along the twist cycle. 
Throughout the parameter sets studied, the overlaps are well controlled, with
$|\mathcal{O}_j(\delta;\theta)|\gtrsim 0.97$ except for the closing link $\theta_{N_\theta-1}\to\theta_0$ given by
\[
\mathcal{O}_{\rm cl}(\delta;\theta)\equiv \langle \Psi_\delta(\theta_{N_\theta-1})\,|\,\Psi_\delta(\theta_{0})\rangle.
\]
This link in the $\theta$-grid gives the minimum overlap, $|\mathcal{O}_{\rm cl}|=\min_j |\mathcal{O}_j|$, and typically lies in the range $0.1$--$0.5$ for the sets of $\delta$, $\lambda$ and $U$ used in this study.
Hence the closing overlap dominantly controls the suppression of $|W(\delta)|$.

\begin{figure}[h]
  \centering
  \includegraphics[width=0.9\columnwidth]{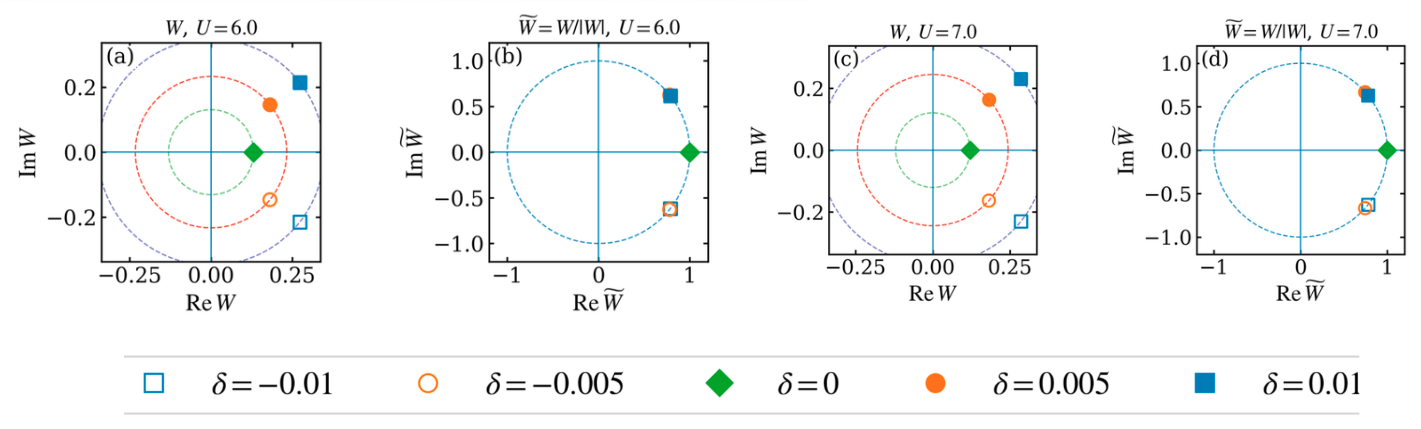}
  \caption{
    Complex-plane plots of the many-body Wilson loop $W(\delta)$ and its normalized version
    $\widetilde W(\delta)\equiv W(\delta)/|W(\delta)|$ for representative $\theta$-grid $N_\theta=48$.
    Panels (a,b): $U=6.0$; panels (c,d): $U=7.0$.
    Markers indicate $\delta\in\{-0.01,-0.005,0,0.005,0.01\}$.
    Marker shape and color encode $|\delta|$ (shared by $\pm\delta$), and filled (open) markers indicate $\delta>0$ ($\delta<0$).
    Dashed circles are shown as guides to the eye.
  }
  \label{fig:W_2x2_raw_unit}
\end{figure}

As also reported in Ref.~\cite{Watanabe2026KS_MB_Berry_SSHH}, we interpret this behavior as a seam-specific suppression associated with the endpoint identification in the uniform gauge, since only the closing link stitches the endpoints $\theta_{N_\theta-1}\approx2\pi$ and $\theta_0=0$.
The absolute value of the closing-link overlap decreases with $N_\theta$ and converges to a finite value ($\sim 0.36$ for $(\delta,\lambda,U)=(0.01,0.67,7.0)$) as $N_\theta$ is increased.
The seam-specific suppression becomes less pronounced as $|\delta|$ increases, consistent with \Fig{fig:W_2x2_raw_unit} where $|W|$ is smallest at $\delta=0$ and increases for $\delta=\pm0.01$.

We also verified that the conjugation identity and the qualitative depinning behavior are insensitive, within the tested range, to standard DMRG convergence controls (bond dimension and sweep number) as well as to system size.

Fig.~\ref{fig:W_2x2_raw_unit} shows the Wilson loop level conjugation identity.
For representative dimerizations, $\delta=\pm 0.005$ and $\delta=\pm 0.01$, we find that the many-body Wilson loop satisfies
\[
W(-\delta)\simeq W(\delta)^*,
\]
within numerical accuracy. (At $\delta = 0$ the Wilson loop is real and satisfies the identity as well.)

\begin{figure}[h]
  \centering
  \includegraphics[width=0.6\columnwidth]{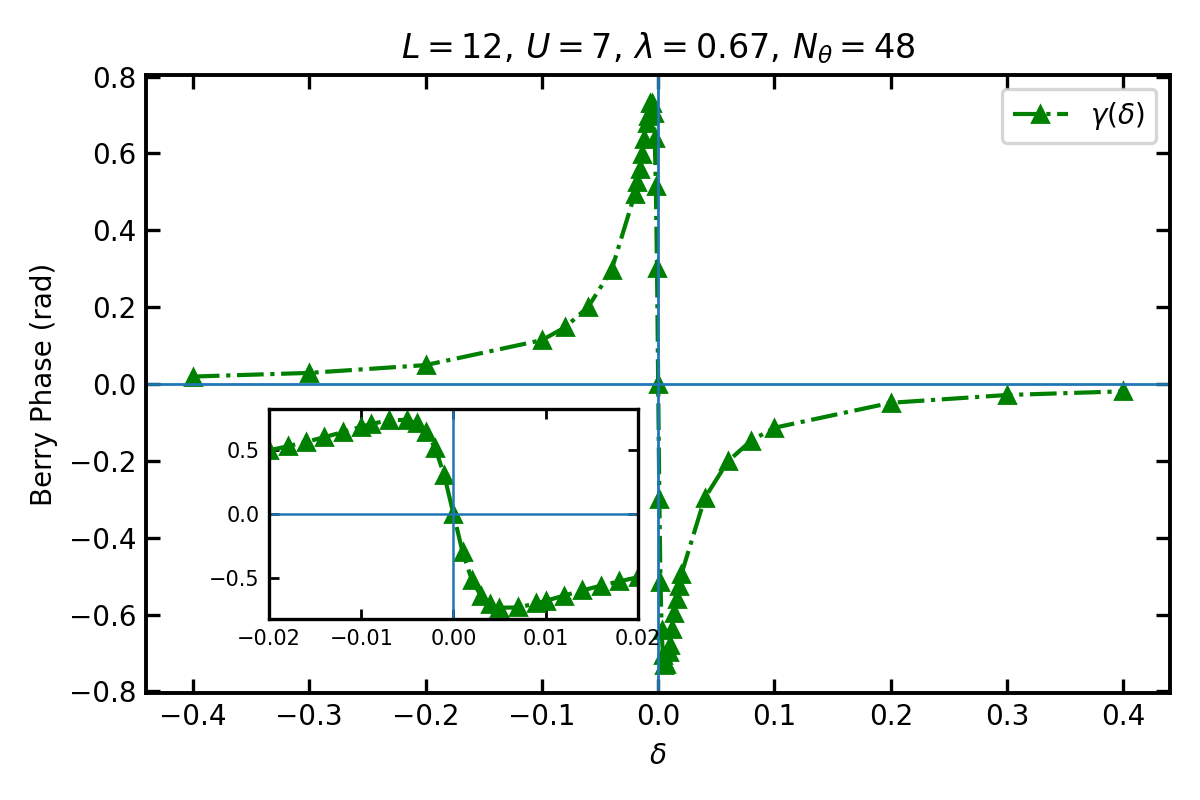}
  \caption{
The Berry phase $\gamma$ is shown as a function of $\delta$ for $U=7.0$, $\lambda=0.67$, and $N_\theta=48$. The inset shows an enlarged view of a part of the plot.
  }
  \label{fig:delta_vs_gamma}
\end{figure}

Since the Berry phase is defined by $\gamma(\delta)\equiv-\arg W(\delta)$, the identity implies $\gamma(-\delta)\simeq -\,\gamma(\delta)$.
Fig.~\ref{fig:delta_vs_gamma} verifies this oddness directly at finite $\delta$ in a regime where $\gamma(\delta)$ varies smoothly.
This shows that the Wilson loop conjugation identity holds without a quantization of the Berry phase.

\begin{figure}[h]
  \centering
  \includegraphics[width=0.7\columnwidth]{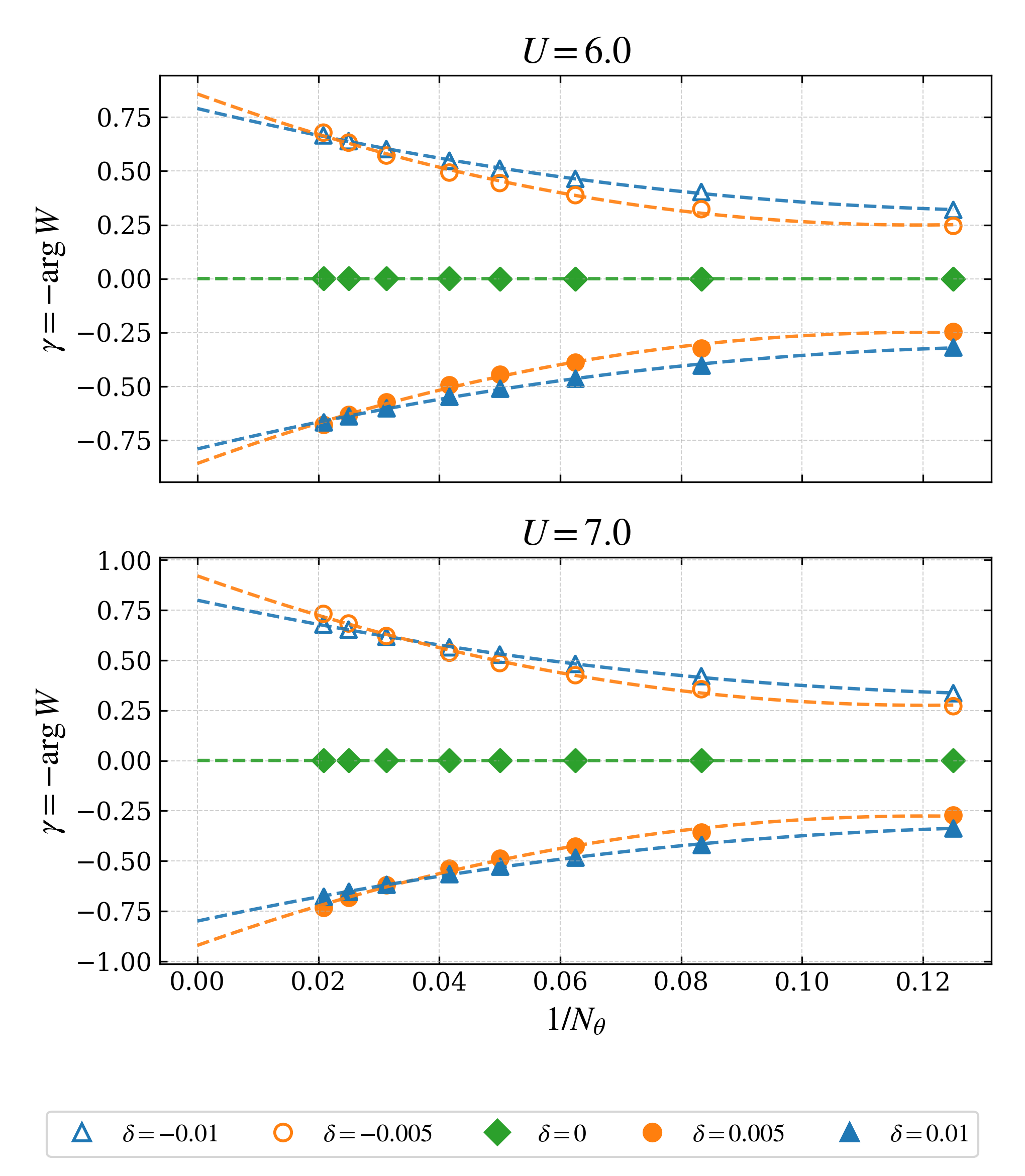}
  \caption{
    Berry phase $\gamma(\delta)\equiv-\arg W(\delta)$ plotted versus $1/N_\theta$.
  Top (bottom) panel: $U=6.0$ ($U=7.0$).
  The data points are for $\delta\in\{-0.01,-0.005,0,0.005,0.01\}$ and $N_\theta\in\{8,12,16,20,24,32,40,48\}$.
  Dashed curves indicate quadratic fits in $1/N_\theta$ to each data set.
  Marker shape and color encode $|\delta|$ (shared by $\pm\delta$), and filled (open) markers indicat\
  e $\delta>0$ ($\delta<0$).
  \label{fig:gamma_fit}
  }
\end{figure}

Fig.~\ref{fig:gamma_fit} summarizes the $\theta$-grid $N_\theta$ dependence of the Berry phase.
We plot $\gamma(\delta)$ evaluated at each $N_\theta$ used in this study, shown as a function of $1/N_\theta$.
To provide a compact parametrization over the tested range, we fit the data as a function of $1/N_\theta$ by
\[
\gamma(\delta;N_\theta)\simeq \gamma_\infty(\delta)+a(\delta)/N_\theta+b(\delta)/N^2_\theta,
\]
where we write the Berry phase with an explicit $N_\theta$ argument to make the twist-grid dependence explicit.
As shown in Fig.~\ref{fig:gamma_fit}, the data are well captured by the fitted curve.
As a representative example, for $U=7.0$ and $\delta=0.01$ we obtain the provisional coefficients
\[
\gamma_\infty(0.01)\simeq -0.800,\ a(0.01)\simeq 6.466,\ b(0.01)\simeq -22.15,
\]
from an unweighted quadratic fit over the tested $N_\theta$ values.
(For this fit, the unweighted RMSE is $\simeq 3.3\times 10^{-3}$ over the tested range.)

Consistent with the conjugation constraint and its Berry phase corollary, the fits for $\gamma(\pm\delta;N_\theta)$ form mirrored curves with opposite curvature.
The quadratic coefficients approximately flip sign under $\delta\to-\delta$; for instance, for $\delta=-0.01$ we find $(\gamma_\infty(-0.01),a(-0.01),b(-0.01))\simeq(0.800,\,-6.464,\,22.14)$.

As an internal large-$N_\theta$ check, for $\delta=0.01$, we also evaluated $\gamma(0.01;N_\theta)$ up to $N_\theta=512$, finding a residual deviation from the quadratic fit of order $|\gamma(0.01;512)-\gamma_\infty(0.01)|\sim 10^{-2}$.
A fully quantitative characterization of $\gamma_\infty(\delta)$ would require a more systematic finite-$N_\theta$ (and finite-$L$) fitting, or alternatively the use of polarization estimators based on the many-body position operator~\cite{Resta1998}.

\section{Discussions}
The key message of this work is that the composite antiunitary mapping associated with the bond-pattern reversal enforces the Wilson loop conjugation identity Eq.~(\ref{eq:W_conj_intro}).
This statement does not rely on any quantization of the Berry phase and remains meaningful in the regime where the Wilson loop is generically complex.

In this work, the bond-pattern reversal is realized by flipping the sign of the dimerization ($\delta\to-\delta$) using the composite antiunitary transformation.

More generally, whenever a nondegenerate flux-threaded ground-state family along the closed $\theta$-cycle, $\{|\Psi_{\mathbf t}(\theta)\rangle\}$, is mapped to $\{|\Psi_{\mathsf P\mathbf t}(\theta)\rangle\}$ by an antiunitary transformation up to a physically irrelevant $\theta$-dependent phase, the corresponding Wilson loop satisfies a conjugation constraint. Here, $\mathsf P$ denotes an appropriate permutation of microscopic parameters, e.g., a bond-pattern reversal of the hopping set $\mathbf t=\{t_{ij}\}$. This generalizes Eq.~(\ref{eq:Xi_maps_psi}) and also applies to continuum models with an analogous loop-reversal mapping; see Supplemental Material~\cite{SM} for a Kronig--Penney example. A related group-theoretical approach can be found in Ref.~\cite{Shiozaki2026Equivariant}.

Beyond its conceptual content, Eq.~(\ref{eq:W_conj_intro}) provides a practical consistency check for numerical evaluations of interacting holonomies.
Deviations from the predicted even/odd structure of $\mathrm{Re}\,W$ and $\mathrm{Im}\,W$ under $\delta\to-\delta$ indicate either explicit breaking of the composite symmetry (e.g., departures from half filling or additional perturbations) or numerical inaccuracies in the overlap-based construction of the Wilson loop.
A stringent internal check is provided by
$
\Delta_W(\delta)\equiv\bigl|W(-\delta)-W(\delta)^{*}\bigr|.
$
One may use the identity to improve the signal in Monte Carlo simulations. For instance, one can sample both $W(\delta)$ and $W(-\delta)$ and average $W(\delta)$ with $W(-\delta)^{*}$. In an ideal case where the two samples are uncorrelated, this averaging reduces the noise by a factor of $\sqrt{2}$.